\documentclass[conference]{IEEEtran}
\usepackage{cite}
\usepackage{amsmath,amssymb,amsfonts}
\usepackage{algorithmic}
\usepackage{graphicx}
\usepackage{textcomp}
\usepackage{xcolor}
\usepackage{comment}
\def\BibTeX{{\rm B\kern-.05em{\sc i\kern-.025em b}\kern-.08em
    T\kern-.1667em\lower.7ex\hbox{E}\kern-.125emX}}
\begin{document}

\title{The Simons Observatory: Magnetic Sensitivity Measurements of Microwave SQUID Multiplexers}

\author{\IEEEauthorblockN{Eve M. Vavagiakis\IEEEauthorrefmark{1},
Zeeshan Ahmed\IEEEauthorrefmark{2}\IEEEauthorrefmark{3}, 
Aamir Ali\IEEEauthorrefmark{4},
Kam Arnold\IEEEauthorrefmark{5}, 
Jason Austermann\IEEEauthorrefmark{6},
Sarah Marie Bruno\IEEEauthorrefmark{7},\\
Steve K. Choi\IEEEauthorrefmark{1}, 
Jake Connors\IEEEauthorrefmark{6}, 
Nicholas Cothard\IEEEauthorrefmark{1}, 
Simon Dicker\IEEEauthorrefmark{8},
Brad Dober\IEEEauthorrefmark{6},
Shannon Duff\IEEEauthorrefmark{6},
Valentina\\ Fanfani\IEEEauthorrefmark{9},
Erin Healy\IEEEauthorrefmark{7}, 
Shawn Henderson\IEEEauthorrefmark{2}\IEEEauthorrefmark{3}, 
Shuay-Pwu Patty Ho\IEEEauthorrefmark{10}, 
Duc-Thuong Hoang\IEEEauthorrefmark{1},
Gene Hilton\IEEEauthorrefmark{6},\\
Johannes Hubmayr\IEEEauthorrefmark{6},
Nicoletta Krachmalnicoff\IEEEauthorrefmark{11}, 
Yaqiong Li\IEEEauthorrefmark{1}, 
John Mates\IEEEauthorrefmark{6}, 
Heather McCarrick\IEEEauthorrefmark{7},
Federico\\ Nati\IEEEauthorrefmark{9},
Michael Niemack\IEEEauthorrefmark{1},
Maximiliano Silva-Feaver\IEEEauthorrefmark{5},
Suzanne Staggs\IEEEauthorrefmark{7}, 
Jason Stevens\IEEEauthorrefmark{1}, 
Mike Vissers\IEEEauthorrefmark{6},
Joel\\ Ullom\IEEEauthorrefmark{6},
Kasey Wagoner\IEEEauthorrefmark{7}, 
Zhilei Xu\IEEEauthorrefmark{8}\IEEEauthorrefmark{12}, 
Ningfeng Zhu\IEEEauthorrefmark{8}}
\IEEEauthorblockA{\IEEEauthorrefmark{1}Cornell University, Ithaca, NY 14853. Email: ev66@cornell.edu\\ 
\IEEEauthorrefmark{2}Kavli Institute for Particle Astrophysics and Cosmology, Menlo Park, CA 94025\\ 
\IEEEauthorrefmark{3}SLAC National Accelerator Laboratory, Menlo Park, CA 94025\\ \IEEEauthorrefmark{4}University of California, Berkeley, CA, USA 94720\\
\IEEEauthorrefmark{5} University of California San Diego, CA, 92093 USA \\
\IEEEauthorrefmark{6}NIST Quantum Sensors Group, 325 Broadway Mailcode 687.08, Boulder, CO, USA 80305\\ 
\IEEEauthorrefmark{7}Princeton University, Princeton, NJ, USA 08544\\
\IEEEauthorrefmark{8}University of Pennsylvania, 209 South 33rd Street, Philadelphia, PA, USA 19104\\
\IEEEauthorrefmark{9}University of Milano - Bicocca, Piazza della Scienza, 3 - 20126, Milano, Italy\\
\IEEEauthorrefmark{10}Stanford University, 450 Serra Mall, Stanford, CA 94305\\
\IEEEauthorrefmark{11}International School for Advanced Studies (SISSA), Via Bonomea 265, 34136, Trieste, Italy\\
\IEEEauthorrefmark{12}{MIT Kavli Institute, Massachusetts Institute of Technology, 77 Massachusetts Avenue, Cambridge, MA, USA 02139}
}}

\maketitle

\begin{abstract}
The Simons Observatory (SO) will be a cosmic microwave background (CMB) survey experiment with three small-aperture telescopes and one large-aperture telescope, which will observe from the Atacama Desert in Chile. In total, SO will field $\sim$70,000 transition-edge sensor (TES) bolometers in six spectral bands centered between 27 and 280 GHz in order to achieve the sensitivity necessary to measure or constrain numerous cosmological quantities. The SO Universal Focal Plane Modules (UFMs) each contain a 150 mm diameter TES detector array, horn or lenslet optical coupling, cold readout components, and magnetic shielding. SO will use a microwave SQUID multiplexing ($\mu$MUX) readout at an initial multiplexing factor of $\sim$1000; the cold (100 mK) readout components are packaged in a $\mu$MUX readout module, which is part of the UFM, and can also be characterized independently. The 100 mK stage TES bolometer arrays and microwave SQUIDs are sensitive to magnetic fields, and their measured response will vary with the degree to which they are magnetically shielded. We present measurements of the magnetic pickup of test microwave SQUID multiplexers as a study of various shielding configurations for the Simons Observatory. We discuss how these measurements motivated the material choice and design of the UFM magnetic shielding.
\end{abstract}


\section{Introduction}

Current and future cosmic microwave background (CMB) experiments rely on superconducting detectors and readout systems which are sensitive to magnetic fields. Experiments like the Simons Observatory \cite{b1}, an array of new CMB telescopes to be located at 5200 m elevation on Cerro Toco in Chile, near the Atacama Cosmology Telescope (ACT) \cite{b2}, CLASS \cite{b3}, and the Simons Array \cite{b4}, use these devices for precision measurements of the microwave sky. CMB map artifacts introduced from devices improperly shielded from scan-synchronous pickup of Earth's magnetic field, radiating half-wave plates, magnetic components inside of cryostats, and other sources could be difficult to remove and jeopardize science goals. 

SO's $\sim$70,000 TES bolometers covering 27--280 GHz will be read out using $\mu$MUX (Microwave SQUID Multiplexing) SQUIDs at an initial multiplexing factor of $\sim$1000. The densely packed 100 mK readout components will be packaged in a $\mu$MUX module within a Universal Focal Plane Module (UFM), which also houses a TES array, optical coupling, and magnetic shielding \cite{b5}. Each UFM will contain $\sim$1800 $\mu$MUX resonators in the 4--8 GHz band, each coupled to a dissipationless radio-frequency
superconducting quantum interference device (RF-SQUID), which is in turn inductively coupled to a TES, and read out using a single pair of coaxial cables \cite{b6}. The resonators and SQUIDs for SO are developed by NIST (National Institute
of Standards and Technology) \cite{b7}.  

The behavior of TESes and SQUIDs under the effect of magnetic fields is not well understood. Theoretical models of the superconducting phase transition of TES bolometers are not sufficiently mature to predict the response to magnetic fields, and the response of SQUID gradiometers changes with each design iteration and is difficult to simulate \cite{b8,b9}. Laboratory measurements of superconducting device magnetic field sensitivity provide valuable data on device response and magnetic shielding requirements, as well as checks on models and simulations. Previous measurements of MoCu and AlMn TES, TDM SQUIDs,
and µMUX SQUIDs have provided comparative estimates of magnetic sensitivities and motivated shielding factors for upcoming experiments that are consistent with shielding factors currently in the field \cite{b10}.

\section{Experimental Setup}

\subsection{512 Box $\mu$MUX SQUID Measurements}

The readout testing was performed using 512 $\mu$MUX resonators and RF-SQUIDs on eight chips installed in the test package referred to as the ``512 box." The NIST $\mu$MUX chips were of microwave SQUID design uMUX100k v1.0, wafer 1, and covered 4--8 GHz. While the $\mu$MUX resonators and RF-SQUIDs may both be sensitive to magnetic fields \cite{b11}, we cannot yet distinguish between the components' behaviors, and thus refer to the magnetic pickup of each combined channel. 

\begin{figure}[ht!]
\centerline{\includegraphics[width=5cm]{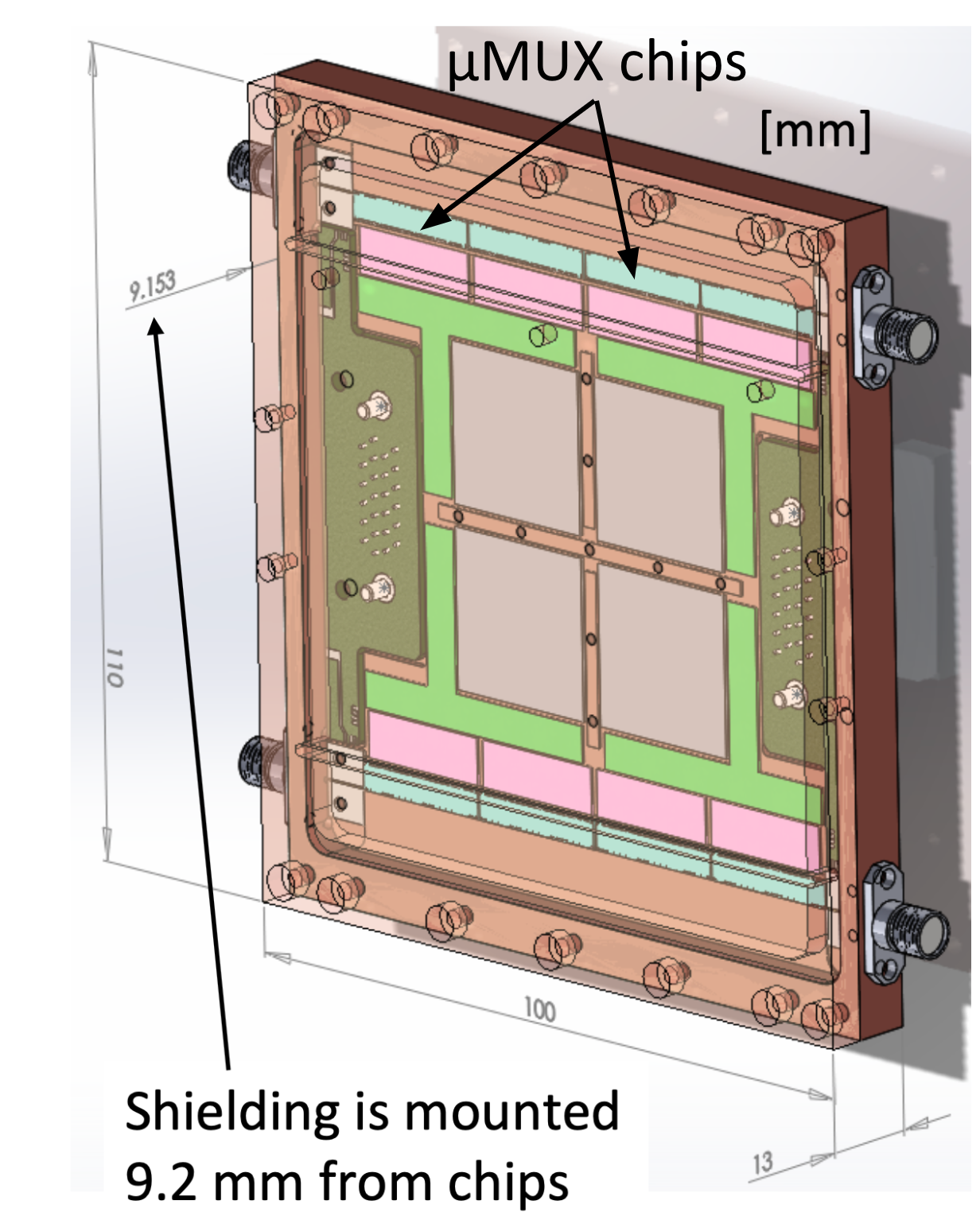}}
\caption{Model of the 512 box test package with eight $\mu$MUX chips (cyan) mounted in a copper box. Shielding material tested was mounted on the copper cover of the box, 9.2 mm from the chip surface. In the case of a sandwich, shielding material was also mounted on the back of the box, equidistant from the chips.}
\label{512diagram}
\end{figure}

\begin{figure}[ht!]
\centerline{\includegraphics[width=9cm]{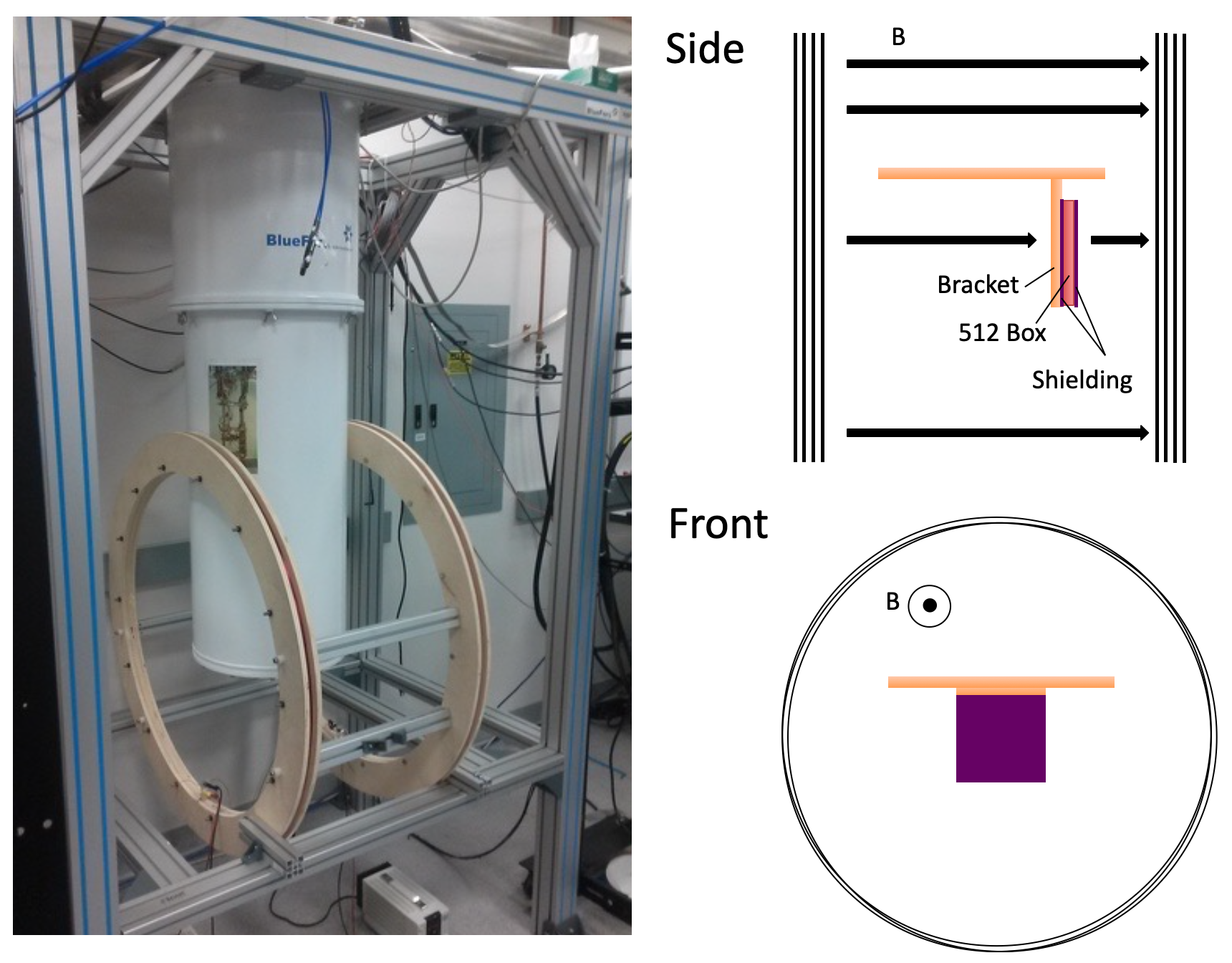}}
\caption{Experimental setup for the 512 box magnetic field testing. The 512 box package was mounted to the 100 mK plate of a Bluefors dilution refrigerator. A set of Helmholtz coils apply DC fields perpendicular to the $\mu$MUX chips, as shown installed around the DR on the left. On the right, a schematic of the DR mixing chamber plate and mounted 512 box test package is shown along with the coils. The ``front" view looks along the z-axis of the coils to the face of the 512 box test package, and the ``side" view shows the x-axis of the coils and the edge of the 512 box package. The orientation of the applied magnetic field is annotated ``B."}
\label{512setup}
\end{figure}

The 512 box (Fig. \ref{512diagram}) was mounted to the cold stage of a Bluefors dilution refrigerator (DR) and cooled to 100 mK (Fig. \ref{512setup}). Any shielding material tested was placed on one or both sides of the test packaging, 9.2 mm from the chips. Any single layer of shielding material was mounted to the side of the 512 box shown in Fig. \ref{512diagram}, while sandwiches of material included a layer alongside the back side in Fig. \ref{512diagram}. 

An external room temperature mu-metal magnetic shield was placed over the DR while the test chips were cooled to avoid trapping magnetic flux in the chips which would degrade their performance. Once at base temperature, this shield was removed, and a set of Helmholtz coils were positioned outside of the vacuum shell of the DR to apply constant DC magnetic fields perpendicular to the plane of the chips (Fig. \ref{512setup}). Fields were not applied parallel to the plane of the chips because they would not couple to the SQUIDs.  No additional magnetic shielding beyond the test pieces near the chips was included. 

\begin{figure}[ht!]
\centerline{\includegraphics[width=9cm]{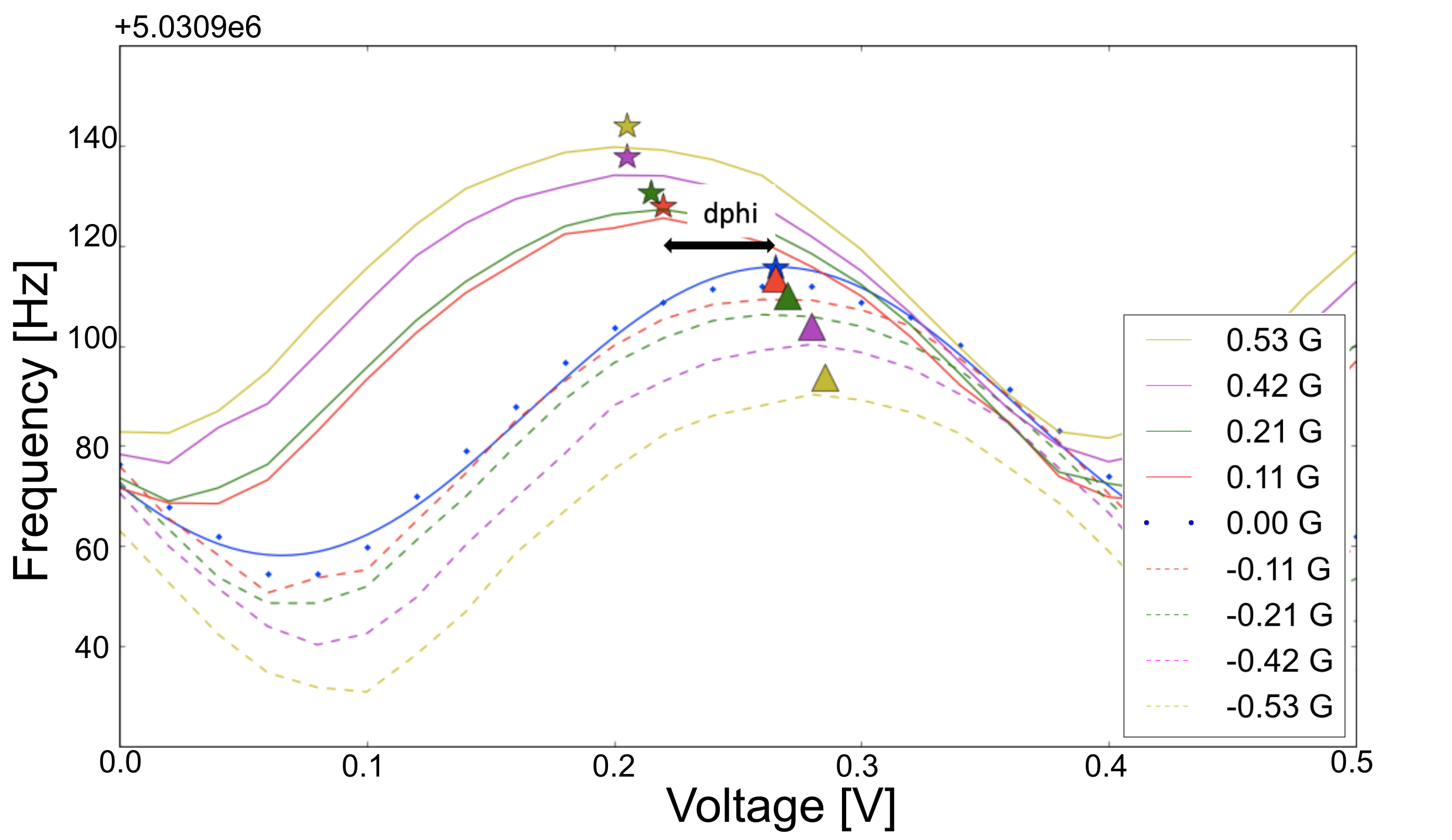}}
\caption{Applied field values shift the SQUID f-$\Phi$ curves as fit to the data acquired with the VNA. An example of the zero applied field data points are plotted in blue, with the sinusoid curve fit blue line overplotted. The offsets in phase (d$\Phi$) are recorded to measure the magnetic pickup in $\Phi_0$/Gauss.}
\label{512shifts}
\end{figure}

87 channels were chosen for analysis, evenly spaced across the 4--8 GHz frequency range. For each data taking run, the same channels were selected for data taking (as chosen by the closest resonance frequency match).  At each applied field value, a vector network analyzer (VNA) traced f-$\Phi$ curves for the chosen channels as the flux through the SQUID ($\Phi$, controlled by voltage applied) was ramped from 0 to 0.5 V in discrete steps of 0.02 V and the minimum frequency of the channel (f) was recorded. Seeking only to measure the phase offsets in $\Phi$, a simple sinusoid was fit to the resulting f-$\Phi$ curves, where one period is $\Phi_0$, the magnetic flux quantum, in this case 0.39 V. The phase offsets of the sinusoid fits from peak-to-peak shifts (d$\Phi$/$\Phi_0$) were recorded for each SQUID (Fig. \ref{512shifts}). The offsets from Earth's magnetic field and the zero-flux phase are accounted for by subtracting the zero applied field d$\Phi$/$\Phi_0$ from all the data points for each channel, such that d$\Phi$/$\Phi_0=0$ for 0 G applied field. The absolute value of these responses is taken so that one linear fit may be performed to the positive and negative shifts. The best fit sensitivity in $\Phi_0$/G for the tested resonances was estimated by fitting a line ($y=\mathrm{m}x$, where $\mathrm{m}=\Phi_0$/G) to the average absolute value of the f-$\Phi$ curve phase shifts as a function of applied magnetic fields. 

\subsection{Magnetic Field Application}

Upon reaching base temperature, the external mu-metal magnetic shield around the DR was removed, and the Helmholtz coils were installed. The highest field value tested (0.525 G) was applied for approximately one minute, and then removed. DC fields were then applied starting at 0 G for a given polarity, and stepped from 0 to 0.525 G in steps of 0.105 G. At each DC field value, f-$\Phi$ curves for each of the resonances were obtained. Fields were applied by setting a power supply to a given current value (i.e. were set suddenly and not ramped up slowly). The field polarity was then reversed, and data were again taken starting from low to high applied field. The field polarities were not tracked between cooldowns, and small relative asymmetries between the two polarities are expected from Earth's magnetic field. 

\begin{figure}[t!]
\centerline{\includegraphics[width=9.2cm]{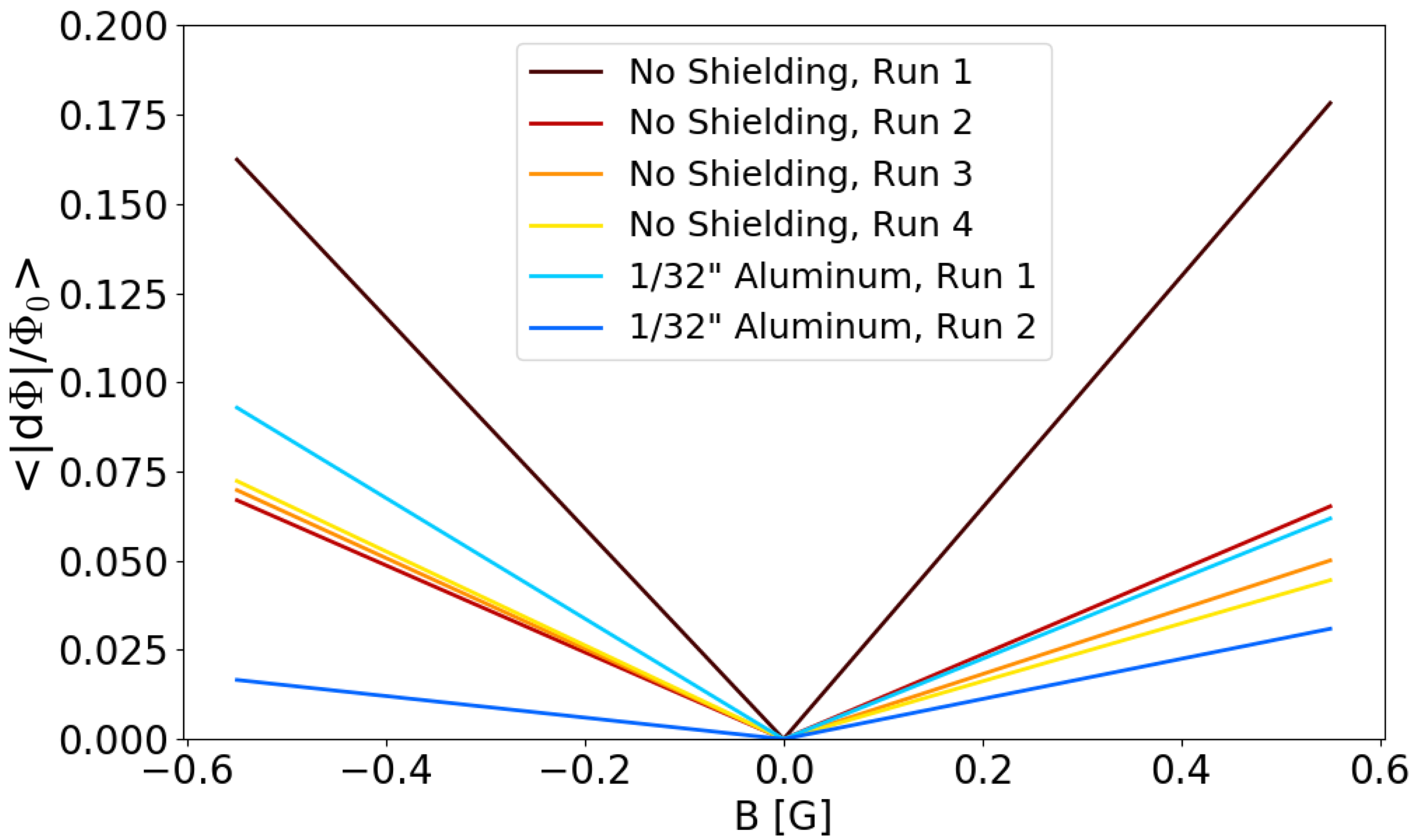}}
\caption{An example of the consequences of ``settling" the $\mu$MUX channels by applying a 0.525 G field for 1 minute before data taking (No Shielding, Runs 2, 3 and 4, 1/32" aluminum, Run 2) versus data taking without first applying the 0.525 G field (No Shielding Run 1, 1/32" aluminum Run 1). Linear fits to the average absolute value of d$\Phi$/$\Phi_0$ for the channels per applied field value are shown for comparison. The average response of the channels decreases after the initial magnetic field is applied.}
\label{512settling}
\end{figure}

\begin{figure}[ht!]
\centerline{\includegraphics[width=9cm]{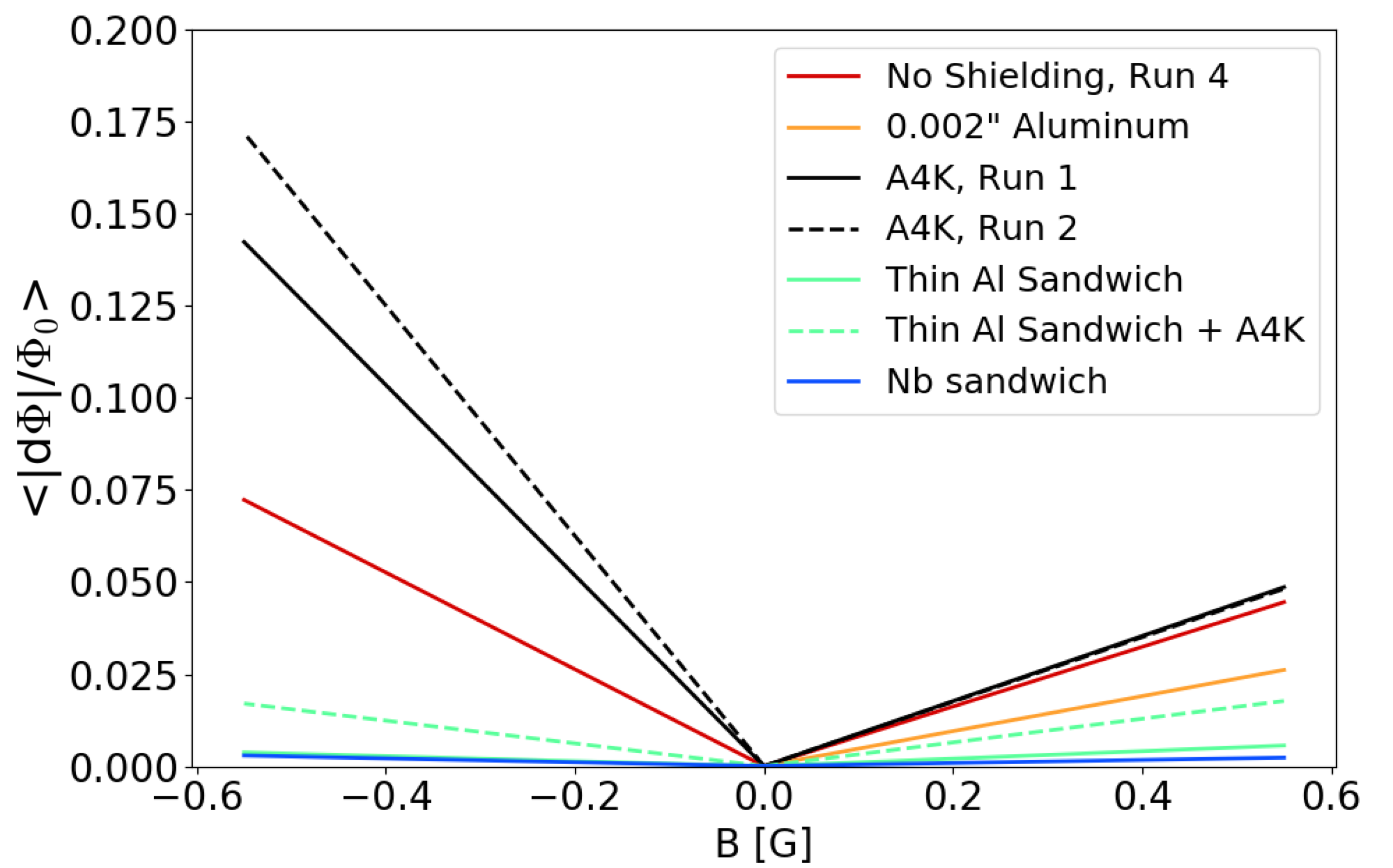}}
\caption{Magnetic pickup experienced by the 512 box channels when shielded by a single piece of A4K (black), a thin (0.002") single piece of Al (blue) or a sandwich of the same (green), the thin Al sandwich plus a piece of A4K (dashed green), or a niobium sandwich (orange), compared to no shielding (magenta). The best shielding configurations were the sandwiches of superconductors, and the worst was the single piece of A4K.}
\label{512materials}
\end{figure}

We found that exposing the chips to the highest value of applied field (0.525 G) had the effect of ``settling" the channels such that their measured sensitivity was lower after this exposure as compared to before. This applied field value is larger than the component of Earth's magnetic field running perpendicular to the surface of the chips, which is estimated to be $\sim$0.05 G. This meant that removing the external magnetic shield and taking measurements starting at 0 G to 0.525 G yields a higher pickup estimate than subsequent repeated set of identical measurements (Fig. \ref{512settling}). This led us to adopt the approach applying the maximum field value (0.525 G) for $\sim$1 minute before taking the data compared in Fig. \ref{512materials}.

No hysteresis was observed when taking data from 0 to 0.525 G in steps of 0.105 G, and then again from 0.525 to 0 G in steps of 0.105 G for the test assembly with a single piece of 0.002" Al. 

\subsection{Shielding Materials}
The magnetic shielding materials tested included two thicknesses (1/32", 0.002") of 6061-T6 aluminum, 0.002" thick Type 2 annealed niobium from Eagle Alloys, and a hexagonal piece (127.20 $\pm$ 0.50 mm corner to corner, $\times$ 1 mm thick) of annealed A4K from Amuneal.

\section{Results}

Tab. \ref{tab1} lists the sensitivity results for the tested magnetic shielding configurations. Average $\Phi_0$/Gauss sensitivities as estimated by linear fits to the average f-$\Phi$ phase offsets for the 512 box $\mu$MUX channels per applied field value are shown for each material configuration. Uncertainties listed come from the linear fits to the average data point per applied field value. A factor of sensitivity reduction from no shielding material is listed.

\begin{figure}[ht!]
\centerline{\includegraphics[width=9cm]{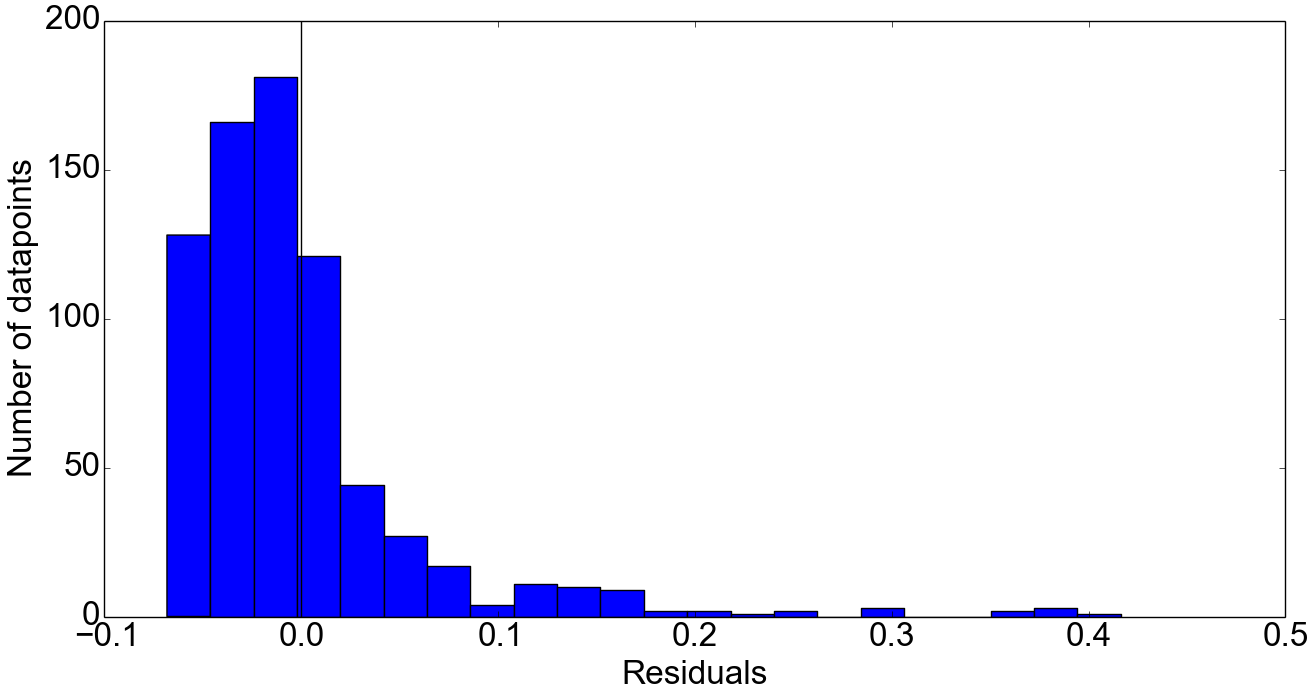}}
\caption{Residuals in $|$d$\Phi|$/$\Phi_0$ from the linear fit to No Shielding Run 2 (Fig. \ref{512settling}). The distribution is typical for all data-taking runs, with most channels showing a lower response than the linear fit to the average, and a small number of channels showing a higher response.}
\label{residuals}
\end{figure}

\begin{table}[htbp]
\caption{Magnetic Shielding Results}
\begin{center}
\begin{tabular}{|c|c|c|}
\hline

\textbf{Material} & \textbf{Sensitivity [$\Phi_0$/G]$^{\mathrm{a}}$}& \textbf{Reduction Factor$^{\mathrm{b}}$} \\
\hline
None & 0.108 $\pm$ 0.015 & 1.0\\
1/32" Al & 0.044 $\pm$ 0.009 & 2.5\\
0.002" Al & 0.053 $\pm$ 0.007 & 2.0\\
A4K & 0.199 $\pm$ 0.091 & 0.5\\
0.002" Al SW$^{\mathrm{c}}$ & 0.009 $\pm$ 0.002 & 12.0\\
0.002" Al SW+A4K$^{\mathrm{d}}$ & 0.033 $\pm$ 0.005 & 3.3\\
0.002" Nb SW$^{\mathrm{e}}$ & 0.005 $\pm$ 0.001 & 21.6\\

\hline

\multicolumn{3}{l}{$^{\mathrm{a}}$Average pickup of the channels for the shielding configuration.}\\
\multicolumn{3}{l}{$^{\mathrm{b}}$Factor by which sensitivity was reduced compared to no shielding.}\\
\multicolumn{3}{l}{$^{\mathrm{c}}$A sandwich of 0.002" Al.}\\
\multicolumn{3}{l}{$^{\mathrm{d}}$A sandwich of 0.002" Al + layer of A4K within the top layer of Al.}\\
\multicolumn{3}{l}{$^{\mathrm{e}}$A sandwich of 0.002" Nb.}
\end{tabular}
\label{tab1}

\end{center}
\end{table}

With no shielding, the 512 box $\mu$MUX channels show a 0.108 $\pm$ 0.015 $\Phi_0$/G shift as estimated using the mean linear fit. This fit is a conservative estimate of the responses, as the distribution of $\Phi_0$/G per channel was not Gaussian, but positively skewed with a lower mode (Fig. \ref{residuals}). The best shielding configurations tested were the sandwiches of 0.002" Al or Nb, with a layer of superconductor on either side of the $\mu$MUX chips, 9.2 mm from the chip surface. These two shielding configurations reduced the shift by a factor of 12.0 or 21.6, respectively. Single layers of Al were the next most successful, providing a 2.0--2.5 factor reduction in field pickup. 

The single layer of annealed A4K yielded higher pickup than no shielding, and introduced an asymmetry in sensitivity depending on field polarity. These results were repeated over multiple data-taking runs. To check for asymmetries introduced by the hexagonal shape of the A4K piece, only channels on the central 512 box chips (which were farthest from the hexagonal edges of the piece) were analyzed. The behavior was the same for these chips alone. When the piece of A4K was included inside the 0.002" sandwich of Al, the shielding was degraded by a factor of 3.6 as compared to the 0.002" sandwich of Al alone.

\section{Conclusion}
NIST test chip $\mu$MUX resonators and RF-SQUIDs for the Simons Observatory were tested for magnetic sensitivity in a variety of magnetic shielding configurations to motivate the design of the UFM shielding. The measured unshielded channel sensitivity of 0.108 $\pm$ 0.015 $\Phi_0$/G is in agreement with the upper bound of 0.3 $\Phi_0$/G previously placed on NIST $\mu$MUX RF-SQUIDs  \cite{b10}. This pickup equates to a 2 $\mu\mathrm{m}^2$ effective cross-section, or one part in $2\times10^4$ of the 40,000  $\mu\mathrm{m}^2$ SQUID area, and is due to imperfect gradiometry, metal symmetry, and uniformity of applied field. Because Earth's field can produce a magnetic flux quantum through just a 40 $\mu\mathrm{m}^2$ area, and the SQUIDs are much larger, gradiometry and magnetic shielding must be combined to provide sufficient insensitivity to Earth's field and other sources while scanning and observing. Layers of superconducting materials 6061-T6 Al and Nb reduce the measured pickup by a factor of 2.0-21.6. This behavior agrees with general expectations as both Type I and II superconductors respond to applied magnetic fields by setting up electric surface currents which cancel out these fields. It is notable that single layers of superconductors provided $\sim$5--11 times less shielding than sandwiches of superconductors. 

The presence of A4K degraded the shielding performance of an Al sandwich by a factor of $\sim$4, and increased the sensitivity of the $\mu$MUX SQUIDs when compared to no shielding at all. The presence of A4K also introduced a large field polarity asymmetry. This motivated the departure from including A4K in the SO UFM packages.

Instead of canceling out applied magnetic fields like superconductors, A4K redirects them. The details of an A4K shielding geometry may thus have profound effects on the material's shielding ability: single flat layers of A4K may not behave like cylindrical shields or boxes, acting to concentrate, distort, and amplify magnetic fields instead of redirecting and attenuating them. It can also be difficult to simulate the behavior of A4K due to numerical issues when modeling high-permeability materials like A4K directly next to zero-permeability materials like superconductors. Our measured results did not agree with our expectations from simulations using ANSYS Maxwell, which estimated that adding a single hexagonally shaped sheet of A4k to the 0.002" Al sandwich would improve, rather than degrade, the shielding performance of the sandwich by a factor of $\sim$4. The disagreement with this result from simulations highlights the importance of laboratory measurements to accurately predict the effectiveness of magnetic shielding materials and geometries.

Along the same lines, our results may not easily extend to real-world shielding designs. Current magnetic shielding tests are being conducted with a more realistic package for an SO $\mu$MUX multiplexer assembly that includes a 6061-T6 Al cover with holes in it for connectors. Introducing holes in this material may degrade its shielding performance. The best performing sandwiches of continuous layers of superconductors may be difficult to implement within the UFM, so further tests are ongoing to define the UFM shielding design. 

Our work has shown that it is difficult to accurately predict the performance of shielding materials in the experimental setting. To properly motivate magnetic shielding designs, laboratory or field tests of realistic materials and geometries are necessary. Such measurements may also improve future models and our understanding of simulations.

\section*{Acknowledgment}

This work was supported in part by a grant from the Simons Foundation (Award No. 457687, B.K.), and by NSF Grant AST-1454881. SKC acknowledges support from the Cornell Presidential Postdoctoral Fellowship and NSF award AST-2001866. ZX is supported by the Gordon and Betty Moore Foundation.

\vspace{12pt}

\end{document}